# Simulation-Based Design of Bicuspidization of the Aortic Valve

Alexander D. Kaiser PhD [1,2], Moussa A. Haidar MD [3], Perry S. Choi MD [3],

Amit Sharir BS [3], Alison L. Marsden PhD [1,2,4,5], Michael R. Ma MD [2,3]

1. Stanford University, Department of Pediatrics, Division of Pediatric Cardiology
2. Stanford University, Cardiovascular Institute
3. Stanford University, Department of Cardiothoracic Surgery, Division of Pediatric Cardiac Surgery
4. Stanford University, Department of Bioengineering
5. Stanford University, Institute for Computational & Mathematical Engineering

## Abstract

Objective: Severe congenital aortic valve pathology in the growing patient remains a challenging clinical scenario. Bicuspidization of the diseased aortic valve has proven to be a promising repair technique with acceptable durability. However, most understanding of the procedure is empirical and retrospective. This work seeks to design the optimal gross morphology associated with surgical bicuspidization with simulations, based on the hypothesis that modifications to the free edge length cause or relieve stenosis.

Methods: Model bicuspid valves were constructed with varying free edge lengths and gross morphology. Fluid-structure interaction simulations were conducted in a single patient-specific model geometry. The models were evaluated for primary targets of stenosis and regurgitation. Secondary targets were assessed and included qualitative hemodynamics, geometric height, effective height, orifice area and billow.

Results: Stenosis decreased with increasing free edge length and was pronounced with free edge length $\leq 1.3$ times the annular diameter $d$. With free edge length $1.5d$ or greater, no stenosis occurred. All models were free of regurgitation. Substantial billow occurred with free edge length $\geq 1.7d$.

Conclusions: Free edge length $\geq 1.5d$ was required to avoid aortic stenosis in simulations. Cases with free edge length $\geq 1.7d$ showed excessive billow and other changes in gross morphology. Cases with free edge length $1.5$-$1.6d$ have a total free edge length approximately equal to the annular circumference and appeared optimal. These effects should be studied in vitro and in animal studies.

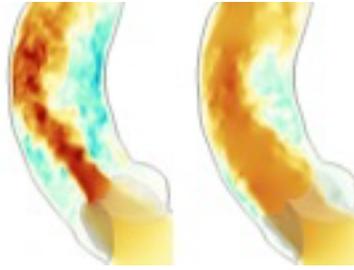

Central Picture Legend 1: In simulations of bicuspidization repair, increasing free edge length reduced stenosis.

Central message: Simulations suggest that total free edge length approximately equal to the annular circumference reduces stenosis in bicuspidization of the aortic valve.

Perspective statement: Bicuspidization of the aortic valve is an effective repair strategy for severe congenital aortic valve pathology. Despite positive outcomes, postoperative stenosis is common. Systematically conducted computer simulations showed that a free edge length of 1.5-1.6 times the annular diameter reduced stenosis. Surgeons should consider increasing free edge length to reduce stenosis and improve outcomes.

Glossary of abbreviations:

| Name | Description |
|---|---|
| $d$ | annular diameter |
| $r$ | annular radius |
| STJ | sinotubular junction |
| $G_h$ | Geometric height |
| $E_h$ | Effective height |
| $E_{h,min}$ | Effective height from minimum height of valve taking billow into account |
| $F$ | Free edge length |

# 1   Introduction

Severe congenital aortic valve pathology in the growing patient remains a challenging clinical scenario—with dysplastic morphology often too complex for common repair strategies and size too small for durable prosthetic options. Over the years, this dilemma has pushed surgeons to devise numerous creative attempts at a durable, surgical solution. First conceived by Schäfers and colleagues, one of these surgical approaches involves bicuspidization of the diseased aortic valve, with particular emphasis on creation of two commissures of equal height.[1] The reconstructed valve resembles a Sievers type 0 pure bicuspid valve.[2] Although originally conceived for unicuspid aortic valve disease—with one fully developed commissure of normal height and two other commissures with rudimentary raphae and lower height—the technique has since been applied in other lesions such as bicuspid, tricuspid, and quadricuspid variants.[3] Despite overall promising results, post-operative stenosis is common, with some studies citing moderate or greater stenosis ($> 20$ mmHg) in up to 75% of patients.[1,3] Further, design principles based on engineering analysis are largely unknown regarding this repair. In this work, simulation-based design tools to optimize the gross morphology of the pure bicuspid aortic valve were developed. The objective was to find a gross morphology that is free of stenosis and regurgitation and create straightforward guidelines that can be tested in future benchtop and animal studies and translated to clinical practice.

The optimal geometry of a pure bicuspid valves appears to be largely unexplored, but related studies have investigated the performance of native bicuspid valves and the typical configuration of a normal trileaflet aortic valve. One study found a rare porcine specimen of a pure bicuspid valve and studied its hemodynamics ex vivo.[4] Most modeling studies focused on bicuspid valves have focused on Sievers type 1 bicuspid valve with a raphe. Emendi et al. investigated the hemodynamics of bicuspid valves in a patient-specific case and compared to clinical imaging.[5] We simulated multiple fusion locations in a single patient-specific geometry and showed that hemodynamics were highly dependent on valve morphology.[6] Lavon et al. studied the effects of cusp angles,[7] and Rego et al modeled strain in patient-specific models including the valve only without fluid.[8] In contrast to the pure bicuspid case, there is substantial literature on the native, presumably optimal, geometry of the healthy trileaflet valve.[9–11] In a previous modeling study, we found that the trileaflet aortic valve was highly sensitive to changes in gross morphology.[12]

In our previous work, methods were developed to construct model valves from as close as possible to first principles[12,13] and used to conduct a controlled comparison of flows associated with different valve phenotypes.[6] A benchtop experiment of flow through a prosthetic valve was conducted and imaged with 4D flow MRI (Four-dimensional flow magnetic resonance imaging). Modeling this experiment and conducting a detailed comparison between experimental and simulation data, revealed excellent qualitative and reasonable quantitative agreement between the two.[14] Thus, our methods are reliable for simulating realistic flow through heart valves. Simulations using related patient-specific modeling methods have been applied to model complex congenital heart disease and impacted clinical practice.[15,16]

In this work, models of the pure bicuspid aortic valve were created and studied with fluid-structure interaction simulations. With fluid-structure interaction simulations, a system of partial differential equations representing the coupled dynamics of the valve and blood was solved to produce detailed predictions of blood velocity and leaflet motion throughout the cardiac cycle. Valve performance was evaluated for stenosis, regurgitation and diastolic gross morphology. We believe this is the first study to simulate flow through the pure bicuspid valve and utilize simulation-based design for bicuspidization of the aortic valve.

## 2 Methods

### 2.1 Simulation methods

The model valves were constructed, as nearly as possible, from first principles via a process referred to as *design-based elasticity*.[12,13] Via the requirement that tension in the leaflets must support a pressure load, an associated system of partial differential equations was derived representing the mechanical equilibrium of the leaflets under pressure. The solution to these equations represented the loaded, closed configuration of the valve. Parameters were tuned in this elasticity problem to design the gross morphology and material properties of the models. Since the valve models were derived from nearly first principles, the emergent material properties mimic those of native tissue and produced effective valve closure in simulations.

Using this process, eight model bicuspid valves were constructed, composed of two equally sized leaflets and two commissures of equal height 180 degrees apart. The planar circle at the base of the annulus is referred to as the virtual basal ring.[17] In this study, the diameter of the entire three dimensional annulus was kept constant at $d = 25$ mm from the virtual basal ring to the sinotubular junction (STJ), including the intercommissural distance. The free edge length at rest varies from approximately $1.1d$, to $1.8d$ in increments of $0.1d$. The model at approximately $1.6d$ has, more precisely, free edge length of $1.57d$, so both free edge lengths add up to approximately $\pi d$, or the circumference of the annulus. As the free edge length increases in the design-based model construction, additional length is added accordingly to the circumferential (commissure to commissure) curves throughout the leaflet, adding area overall to the leaflet including the leaflet belly. An additional set of models were constructed that all have free edge length of approximately $1.6d$ with less leaflet area or "bowl" in the leaflet belly, less leaflet height and both less bowl and height (see Supplemental Information). The leaflets were attached to a rigid scaffold that supports the valve at both commissures. The diameter of the scaffold is constant throughout the annulus from the virtual basal ring to the STJ, which forces the intercommissural distance to be equal to the annular diameter. One commissure was aligned with the patient's left- and non-coronary commissure, since this commissure is most frequently normal in patients with a unicuspid valve and left in position.[2,3] The opposing commissure is supported by the scaffold, which covers any residual space between the valve and sinus.

A patient-specific model aorta geometry was constructed in which to mount the model valves. An anonymized CT scan of a patient with no known aortic or valvular disease was retrospectively acquired. The Institutional Review Board (IRB) of the Stanford University approved the study protocol and publication of data. Patient written consent for the publication of the study data was waived by the IRB for use of anonymized, retrospectively acquired data (#39377, 6/27/2023). The aortic geometry was manually segmented from this CT scan using SimVascular.[18] This patient had approximately equal diameter of the aortic annulus from the virtual basal ring to the STJ.

Fluid-structure interaction simulations were performed with the immersed boundary method, a numerical method that is well-suited to simulations involving heart valves,[19,20] using the open-source solver IBAMR (immersed boundary adaptive mesh refinement).[21] Simulations were driven by prescribing pressures at the inlet, approximately at the left ventricular outflow tract, and the outlet, in the distal ascending aorta. More details are described in the supplemental information.

## 2.2   Evaluation metrics

Scalar metrics were computed to evaluate valve performance. In systole, orifice area of the open valve, the mean pressure gradient in mid systole and the stroke volume were computed. In diastole, metrics were computed based on gross morphology (Figure 1). The geometric height $G_h$ represented the three-dimensional length of the midline curve of the leaflet from annulus to free edge. The quantity $G_h/r$ is the geometric height normalized to the radius of the annulus. The billow height measured the lowest point of the leaflet in the direction normal to the virtual basal ring; if this quantity is positive there is no billow. The effective height $E_h$ measured the distance from the virtual basal ring in the normal direction to the center of the free edge of the leaflet. The effective height measured from the minimum height of the valve is denoted $E_{h,min}$ and takes billow into account. The quantities $E_h/G_h$, $E_{h,min}/G_h$ represent the ratio of effective height over geometric height and the ratio of effective height considering billow over geometric height. The free edge length $F$ represented the three-dimensional length of the leaflet as it runs from commissure to commissure, which was computed on the aortic side of the free edge. The quantity $F/d$ represented the free edge length normalized by the valve diameter.

## 3   Results

Fluid-structure interaction simulations were performed on cases with varying free edge lengths. Cases with free edge rest length ≥1.5$d$ were free of any stenosis. Cases with free edge rest length ≤1.3$d$ showed increasing stenosis with decreasing free edge length. The 1.4$d$ case showed marginally higher pressure gradient than the larger cases. The cases with free edge rest lengths ≥1.7$d$ showed good systolic performance, but excessive billow during diastole.

The dynamics of blood velocity and valve motion on the 1.57$d$ case through the cardiac cycle appear qualitatively excellent, with no apparent stenosis or regurgitation (Figure 2, Video 1, Video 2). The top row shows the vertical component of velocity on a slice through the center of the virtual basal ring. The middle row shows the axial component of velocity on five slices that are approximately normal to the axis of the vessel. The slices are located at the virtual basal ring, the STJ, and the bottom, middle and distal portions of the ascending aorta. The bottom row shows the valve from above along with the velocity normal to the virtual basal ring. If the leaflets billowed, the velocity at the virtual basal ring appeared above the leaflets in the visualization, serving as a visual queue for billow.

The pressure and flow waveforms during the second cardiac cycle on the 1.57d case suggested the valve was free of dysfunction (Figure 3). The flow rate showed an oscillation that rapidly decayed in amplitude then remained nearly zero through diastole, then rose rapidly in systole. The waveforms indicated that, under physiological pressures, the valve opens freely and closes reliably.

The blood velocity and valve position in mid-systole varied substantially across cases (Figure 4). The cases with shorter free edge lengths showed stenosis via a visibly narrowed valve orifice and jet of forward flow. The 1.1$d$ case appeared highly stenotic, with a large fraction of the orifice obstructed by the leaflets. At the STJ, the narrowed jet of forward flow is surrounded by relatively static flow. Just above, regions of backflow surrounded the jet. In the distal ascending aorta, the jet impacted the greater curvature of the vessel, and a region of reverse flow was apparent on inner side or lesser curvature. The narrowed jet thus caused local regions of retrograde flow. The 1.2$d$ case showed similar features to the 1.1$d$ case but was modestly less stenotic with a slightly wider jet. The $1.3d$ case was similar to the $1.2d$ case and was omitted for clarity. The $1.4d$ case showed a subtly narrower jet of forward flow and larger areas of backflow than the cases with more free edge rest length.

The cases with more free edge length showed uninhibited forward flow and no apparent stenosis. The $1.57d$ case, which has a free edge rest length approximately equal to the circumference of the annulus, showed a wide opening and jet of forward flow and negligible backflow. The $1.5d$ case appeared nearly identical to the $1.57d$ case and was omitted. The $1.8d$ case had similar unrestricted forward flow to the $1.57d$ case but appeared to have "excess" leaflet material in the sinus, with rippling or bunching of the leaflets visible. The $1.7d$ case was showed similar excess material and was omitted.

Billow increased with increasing free edge length in mid-diastole (Figure 5). In all cases, there was no visible regurgitation. The distal ascending aorta and middle panel were omitted, as blood velocity is nearly zero. One the $1.1d$ case, the top view showed a fairly straight line of coaptation and small regions of billow on each leaflet belly. On the $1.2d$ case, the free edge showed a subtly larger region of billow. The $1.4d$ case showed more redundancy on the free edge length, with a slight ripple forming. On the $1.57d$ case, the leaflets are visible below the virtual basal ring, and a portion of the leaflet belly dips below the ring. On one leaflet, a small fold of redundant material was present on the free edge. On the $1.8d$ case, the majority of the leaflet belly billowed below the virtual basal ring. Even the center of the free edge dips slightly below the plane of the virtual basal ring, making for a negative effective height. The $1.3d$, $1.5d$ and $1.7d$ cases were similar to the cases of comparable free edge lengths and omitted.

The pressure gradient decreased, and flow rate increased monotonically with increasing free edge length up to $1.5d$, above which there was little difference in pressure and flow rate (Figure 6). The $1.1d$ case is highly stenotic, with a sustained pressure gradient over 30 mmHg and a correspondingly low flow rate. The $1.2d$ is also stenotic, with peak pressure gradient of over 20 mmHg. The $1.3d$ case shows a slightly lower pressure gradient and modest loss of flow rate. The $1.4d$ case shows a subtly higher pressure gradient and lower flow rate than subsequent cases, indicating that there is still insufficient free edge length to minimize the pressure gradient. The $1.5d$ to $1.8d$ cases all show similar pressure gradients and flow rates, indicating that the free edge length is sufficient to open widely and leave the annulus free of obstruction.

Scalar metrics varied widely in both systole and diastole across cases (Table 1). The geometric heights and nondimensional geometric heights are relatively consistent throughout the models, by design. The billow height increased monotonically with free edge length, with substantial billow on the $1.7d$ and $1.8d$ cases. The effective height varies widely, from > 1.1 cm to <0 cm, meaning that part of the free edge in the most billowed case was below the virtual basal ring. The effective height from minimum decreases similarly but is, by definition, larger. The effective height to geometric height ratio was 0.5 in the $1.1d$ case, ~0.3 for the $1.4 - 1.57d$ cases, and approximately zero for the most billowed cases. Taking billow into account, the ratio $E_{h,min}/G_h$ was $\geq 0.39$ for the 1.1-1.57 cases, but <0.3 in the 1.7 and $1.8d$ cases. The nondimensional free edge length $F/d$ in all cases was slightly larger than the free edge rest length for each case, as expected since the free edge was loaded. In systole, the orifice area increased, the pressure gradient decreased, and the stroke volume increased with increasing free edge length up to the $1.4d$ case. The 1.5-1.8$d$ cases showed approximately constant area, pressure gradient and stroke volume. This behavior suggests that adding additional free edge rest length beyond $1.5d$ had little effect on the systolic performance of the valve.

Models with consistent free edge length and altered gross morphology are discussed in the supplemental information.

## 4    Discussion

Fluid-structure interaction simulations were conducted to optimize the gross morphology of bicuspidization of the aortic valve. Configurations with free edge rest length 1.3 times the annular diameter or less showed clinically relevant stenosis with mean pressure gradients exceeding 12 mmHg. Cases with free edge rest length 1.57 times the diameter or greater showed sustained pressure gradients below 8 mmHg. Increasing the free edge length further did not reduce this gradient. The case with free edge rest length 1.4 times the diameter had a slightly higher pressure gradient than the cases with free edge rest length 1.57 times the diameter or greater. Configurations that avoided stenosis had free edge lengths that were relatively large, meaning the leaflets also had large surface area. This large surface area led to leaflet billow and decreased effective height, though without apparent regurgitation.

Based on these results, we propose that free edge rest length of approximately $1.57d$ on each leaflet is optimal for reducing stenosis. This case had no apparent stenosis and unrestricted forward flow, but less billow and excess material than cases with more free edge length.

The following calculation explains some of the challenges of constructing a bicuspid valve with two equal leaflets. When the closed geometry is projected onto a two-dimensional plane, the projected free edge length is approximately $2d$. To open fully and avoid any obstruction to forward flow or stenosis, the free edge length must be approximately the same length as the circumference of the aortic annulus, or $\pi d$ or just over $3d$. This calculation suggests an optimal total free edge length slightly exceeding $2d$ in diastole, accounting for the central coaptation point being lower than the commissures. The optimal length in systole, however, is slightly larger than $3d$. Further, in systole the leaflets are less loaded and thus less stretched and smaller. Therefor additional redundant length of order $d$ in the diastolic configuration is necessary for a non-restrictive systolic configuration. Our predicted optimal free edge length of $1.57d$ for each leaflet and $3.2d$ in total thus provides sufficient length for good opening, while showing only modest billow compared to the 1.7 and $1.8d$ cases.

Similar estimates for a healthy, normal, trileaflet valve starkly contrast with these results. Again projecting onto a two-dimensional plane, each leaflet has free edge length $d$ so the total projected free edge length is $3d$, and the three dimensional length is just over $3d$. Then with a small relative change in length, the leaflets can deform from the closed configuration to an open configuration with free edge length approximately equal to the circumference of the annulus, which is $\pi d$ or just over $3d$. This suggests a much smaller relative change in leaflet lengths through the cardiac cycle is necessary with normal, trileaflet anatomy.

In this study, it was assumed that the annular diameter is constant from the virtual basal ring to the STJ. In cases where the diameters differ, the STJ diameter, which is equal to the intercommissural distance, is likely the relevant diameter operatively. Since the STJ can be more easily remodeled operatively, however, using the virtual basal ring diameter and resizing the STJ accordingly may be an attractive clinical option. With a constant leaflet geometry on a trileaflet case, a modeling study showed substantial differences in coaptation with changes in the virtual basal ring diameter while keeping the STJ diameter fixed.[22] Thus, a systematic comparison of optimal performance with variations in diameter at the virtual basal ring and STJ would be highly relevant for future work.

A number of minor limitations were present in this study. Compensatory measures for left ventricular pressure were not considered, but if they were included, the increases in flow rate would have caused even more pressure gradient to occur in the stenotic cases. Leaflet material properties were modeled approximately as healthy native tissue, whereas surgical reconstruction would include pericardial or synthetic material. Suture lines may further influence valve kinematics. The model valves were all mounted to a stiff scaffold (like a virtual sewing ring) which rigidly holds the commissures precisely in place 180 degrees across the annulus, which may not be possible to achieve depending on the patient's anatomy.

We believe the general trends presented in this work are widely applicable to any surgeon interested in optimizing bicuspidization of the aortic valve. Further laboratory, animal and human studies should be conducted to validate these predictions and apply them to clinical practice.

## 5    Conclusions

Simulations showed that stenosis decreased with increasing free edge length, and that with free edge lengths ≥1.5 times the annular diameter, no stenosis occurred. The optimal configurations showed substantial extra, redundant leaflet material and free edge length when closed, resulting in leaflet billow and the appearance of excessive free edge length in diastole. Despite an imperfect appearance when closed, these valves functioned well, without stenosis or regurgitation. In contrast, configurations with a shorter free edge length had less billow and a straighter, more "crisp" coaptation line, but stenosis occurred during systole. These models, despite a subjectively better appearance in diastole, showed dysfunction in systole.

While the optimal repair may depend on the individual anatomy, the general trend predicted in this work is clear: With inadequate free edge length, bicuspid valves showed stenosis, and augmenting free edge length resolved the stenosis.

Our results thus suggest that surgeons should consider including additional free edge length to avoid stenosis in surgical bicuspidization of the aortic valve.

## 6    Acknowledgements

ADK was funded in part by a grant from the American Heart Association #24CDA1272816. PSC was funded in part by a grant from the National Institutes of Health #R38HL143615. Figure 1 was illustrated by Michael Leonard of University of the Pacific. Computing for this project was performed on the Stanford University's Sherlock cluster with assistance from the Stanford Research Computing Center. Simulations were performed using the open-source solver package IBAMR, https://ibamr.github.io.

**Supplemental information**

# 8 Methods

Additional details on the methods are provided here.

## 8.1 Construction of the model valves

The model valves were constructed via *design-based elasticity,* in which designed the closed configuration of the valve was designed via tuning parameters in a nonlinear partial differential equation.[12,13] This differential equation was derived from the requirement that tension in the valve leaflets supports a constant pressure load. The solution represents the predicted, loaded, closed configuration of the valve. From the closed configuration, a reference configuration and constitutive law for the leaflets was derived. Experimentally measured values of stretch (ratio of current length to reference length) were then prescribed in the radial and circumferential directions, then derived reference lengths for the discretized model from these stretch ratios.[23] The shape of the stretch/tension response was exponential and zero at a stretch ratio of one; the exponential rates were selected from experimental data.[24] The stiffnesses of the exponential responses locally were scaled to have tension needed by the predicted loaded configuration at the experimentally measured stretches. This completed the derivation of the reference configuration and constitutive law from the model valve.

To obtain an initial configuration suitable for fluid-structure interaction simulations, an ideal free edge configuration to reduce residual stress in the leaflets was computed algorithmically with an iterative search for an ideal free edge position. This method searched for uniform stretch circumferentially and a specified pre-stretch along each (mathematical) fiber in the radial direction. First, a constant pre-stretch was prescribed to the radial direction fibers based on their predicted loaded stretch. An arbitrary curve of the form $a\sin^2\theta$ was computed from each of the commissures, where the optimal value of $a$ was initially unknown. Each free edge point was then placed on a ray from the annulus to this curve at the prescribed distance from the annulus. The value of $a$ was then optimized with Matlab's fsolve such the length of the resulting curve was approximately equal to the free edge rest length. This produced a configuration with the desired radial stretches and total free edge length, but without locally uniform stretch along the free edge. The following two steps were then repeated until the root mean squared difference of stretch along the free edge was less than $10^{-10}$ or after 2000 iterations. The free edge points were then distributed on a linear interpolant between the current points, spaced as a fraction of the rest length of each edge in the discretized model. To re-obtain the prescribed radial stretch, each free edge point was again placed on a ray from the annulus at the required distance. In most cases, this process converged resulted in uniform stretches circumferentially on the free edge, the prescribed radial pre-stretch. In some cases, the valve lacked sufficient circumferential or radial length for this process to converge and a nonuniform pattern of stretch occurred on the free edge. Nonetheless, running this algorithm still reduced the residual stretch and tension in the leaflets and it was applied it to all cases. This location was then fixed as a Dirichlet boundary condition and a problem of mechanical equilibrium was solved with the reference configuration and constitutive law that was just derived with zero pressure loading. This process resulted in lower stretches throughout the valve. The leaflets are set to a thickness of 0.44 mm.[25] This configuration was then used as initial conditions for fluid-structure interaction simulations.

## 8.2 Construction of the model aorta

To construct the model aorta with SimVascular, a path line was selected that represents the approximate centerline of the vessel. SimVascular then computed image slices normal to this path. The outer boundary of the lumen was manually segmented on these two dimensional from the left ventricular outflow tract through the distal ascending aorta. These two-dimensional segmentations were lofted into a three-dimensional surface representation, and a triangular mesh for this geometry was computed and exported for use in fluid-structure interaction simulations.

Flow extenders were added to the inlet and outlet of this geometry. The extender at the outlet was cylindrical and 1 cm long. The extender at the inlet was 3 cm in length, the first 1.5 cm constricts smoothly by 5 mm and is followed by 1.5 cm with a constant cross-sectional profile. Introducing this smooth constriction in the flow extender ensured numerical stability at the inlet during forward flow, and the flow averaging force of our previous work was omitted.[6] The position was prescribed to be nearly rigid via stiff linear springs on the faces of the mesh and target points, stiff linear springs of zero rest length tied to a fixed location.

## 8.3 Fluid-Structure Interaction

The immersed boundary method uses a regular, Cartesian mesh to model the fluid, blood, whereas the structure, which includes the leaflet, scaffold on which the leaflets are mounted and model aorta, is modeled as a continuum of mathematical fibers, which exert force along their axes. The fluid domain mesh occupied the entire region of interest, and the structure mesh occupied a proper subset of the fluid domain. At the interface of the fluid and structure, the meshes are not required to conform to each other. Coupling between the fluid and structure is handled via convolution with a discrete approximation to the Dirac delta function. These convolutions ensure that the structure moves with the fluid velocity, and force from the structure is appropriately distributed to the fluid. The use of non-conforming meshes allows for FSI simulations with large elastic deformations of valve leaflets, rapid changes in the fluid domain geometry and changes in the fluid domain topology. Due to the diffuse interface coupling scheme, the structures such as leaflets interact when they are a small but finite difference from each other, and no additional contact forces were required.[26]

All simulations were run with the open-source software package IBAMR (Immersed Boundary Adaptive Mesh Refinement).[21] The spatial resolution was set to 0.5 mm for all simulations. In a convergence study in previous work, little difference was found between results with this resolution and 0.375 mm and concluded 0.5 mm was sufficient resolution.[7] The time step was set to either $6 \cdot 10^{-6}$ or $5 \cdot 10^{-6}$ s, the minimum that was found to be stable for each specific case. In previous work, little difference was found between the second cycle and third and fourth cycles. Thus, all cases were run for two cardiac cycles of 0.8 s duration, or 75 beats per minute. The first cycle was discarded due to initialization effects, and present results from the second cycle in this work. All simulations were run on 48 Intel Xeon Gold 5118 cores across two nodes with a 2.30 GHz clock speed via the Sherlock cluster at Stanford University.

## 8.4 Evaluation Metrics

To evaluate regurgitation, the flow rates through the artery were computed by directly integrating the fluid velocity field at the inlet within IBAMR. To evaluate stenosis, the mean pressure was computed directly from the simulated fluid field. Using Paraview, the pressure was integrated over two-dimensional slices in left ventricular outflow tract and the STJ; the gradient reported is the difference of these two quantities. (Note that this is formally a pressure difference rather than a pressure gradient, but the term pressure gradient is used, as is standard in clinical practice.)

The orifice area during systole was computed as follows. A rigid body motion that maps the virtual basal ring to the unit circle was applied, then each point of the valve was projected onto the plane of the virtual basal ring. Each point was then reflected over the unit circle and computed the outer boundary of these reflected points via Matlab's boundary function. This produced a tight bounding polygon to this set of points. Using the original, rather than reflected, position of the points, this area of this polygon is approximately equal to the area internal to the valve.

## 9    Results

Three additional cases were run all with free edge length of approximately $1.6d$ in which the gross morphology varied in other ways. One case had less area or "bowl shape" to the leaflets, which created a more tube-like geometry with less leaflet material to billow. One case had less leaflet height, and one had both less area and less height. These models were constructed to investigate the substantial variability in gross morphology that may occur even with consistent free edge length.

A visualization of four cases in systole and diastole shows subtle but substantial differences between cases, where the "basic" version is identical to the $1.57d$ case used elsewhere in the paper (Supplementary Figure 1). With area removed, the systolic flow appeared qualitatively similar to the basic case. The systolic configuration of the leaflet appeared straighter and more cylindrical. In diastole, the valve appeared to be coapted below the free edge, and the free edge shows redundant length in this configuration. The case with height removed showed qualitatively similar flows and valve position to the basic case in both systole and diastole. The case with both height and area removed shows slightly stronger recirculation and cylinder-like shape during systole. In diastole, this case showed less billow height and billowed over a smaller portion of the leaflets.

Despite relatively similar performance, the scalar metrics still showed variation between cases (Table 2). The cases with reduced area showed modestly reduced orifice area and stroke volume and modestly increased pressure gradient, worse in the case that also had reduced leaflet height. These cases showed pressure gradients and stroke volume similar to that of the $1.4d$ case, despite having longer free edge lengths. In diastole, the cases with reduced height showed lower geometric height, as expected. Compared to cases with similar leaflet heights, the cases with reduced area showed reduced billow, increased effective height, effective height from minimum, effective height to geometric height ratio and effective height from minimum to geometric height ratio. Thus, despite worse systolic performance, the cases with reduced area appeared to have a better diastolic configuration with less billow and better effective height.

These cases illustrate that the three-dimensional geometry of the leaflets influences billow and forward flow behaviors, even when two fundamental quantities that control leaflet size – free edge rest length and leaflet height – remain the same. Further, there appears to be a consistent tradeoff between systolic performance and the presence of billow and other poor diastolic behaviors. Experimental studies must be conducted to understand the clinical consequences of such changes.

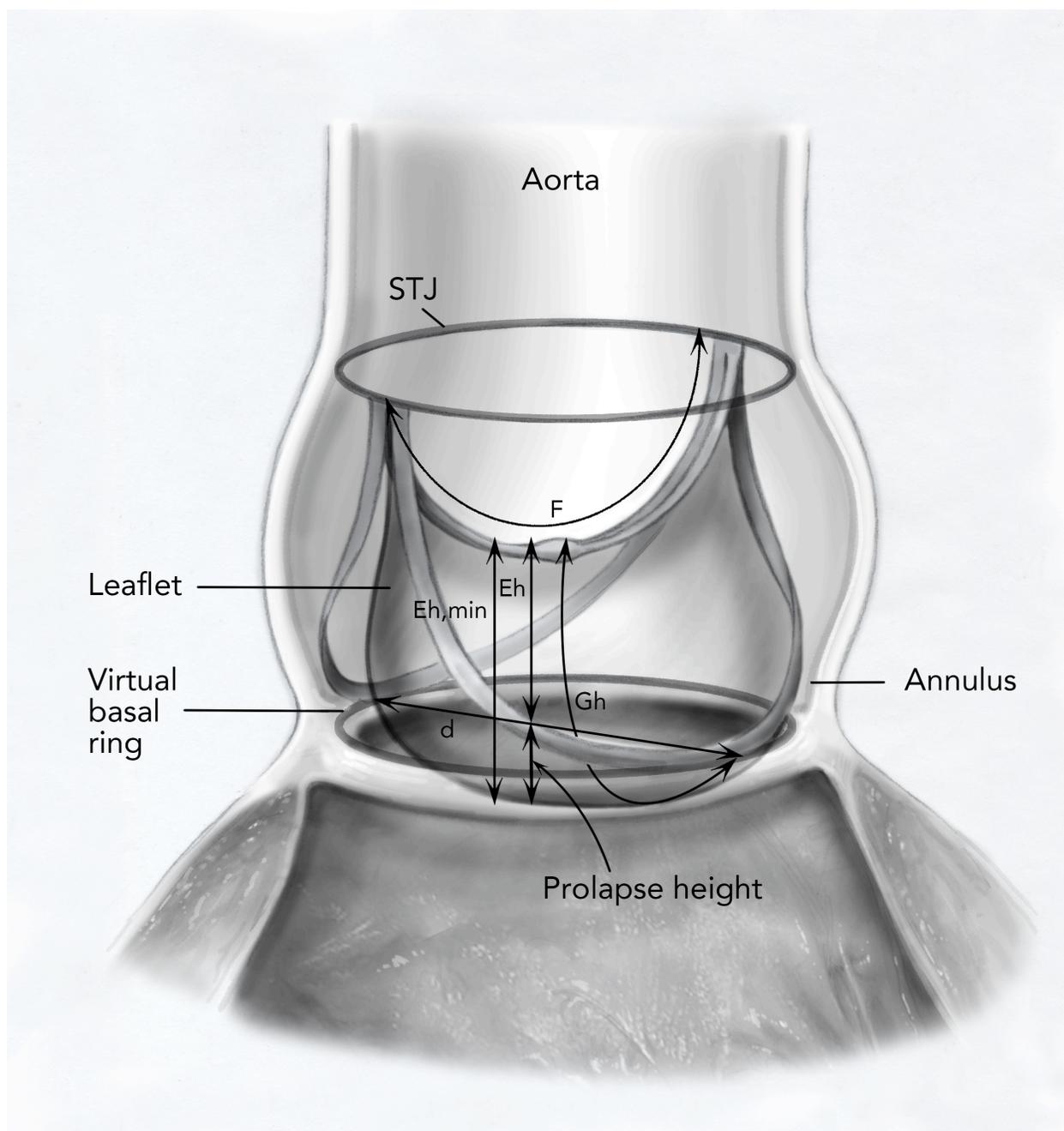

Figure 1: Schematic showing gross morphology.

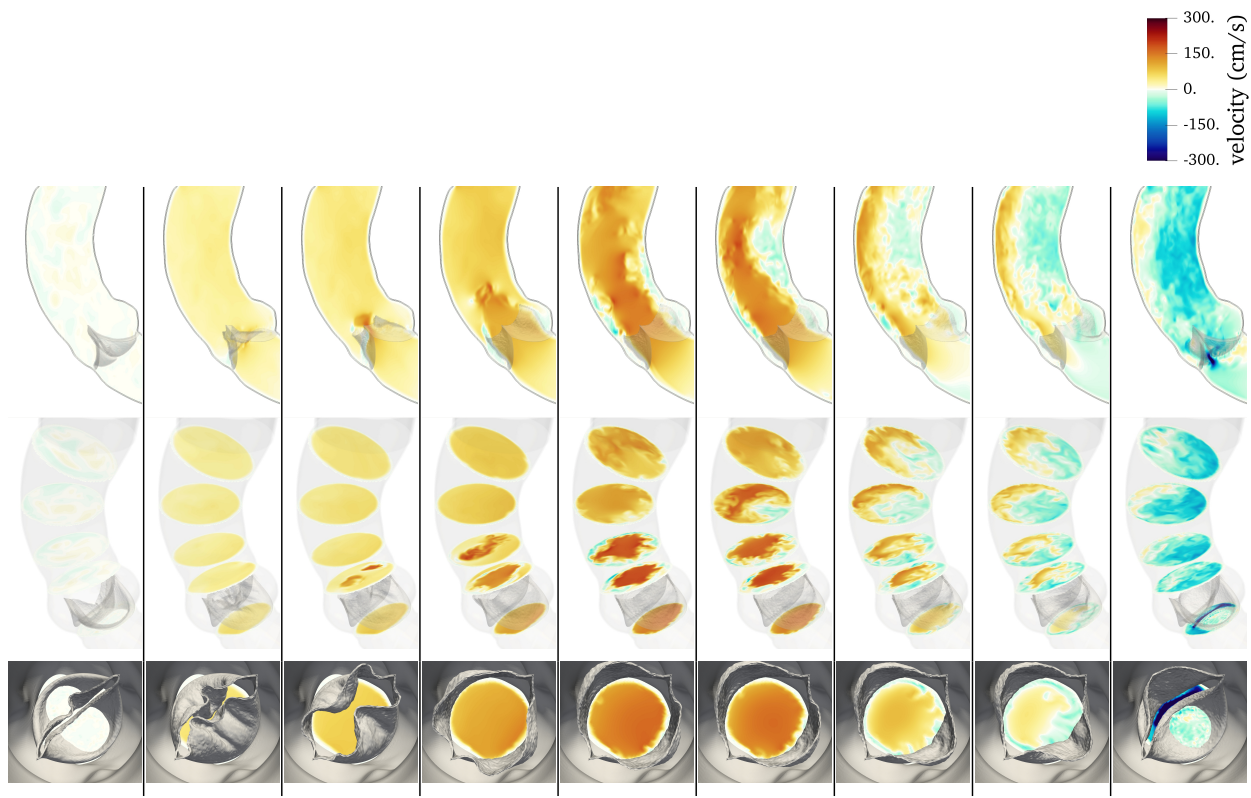

Figure 2: Flows through the cardiac cycle, 1.57d case

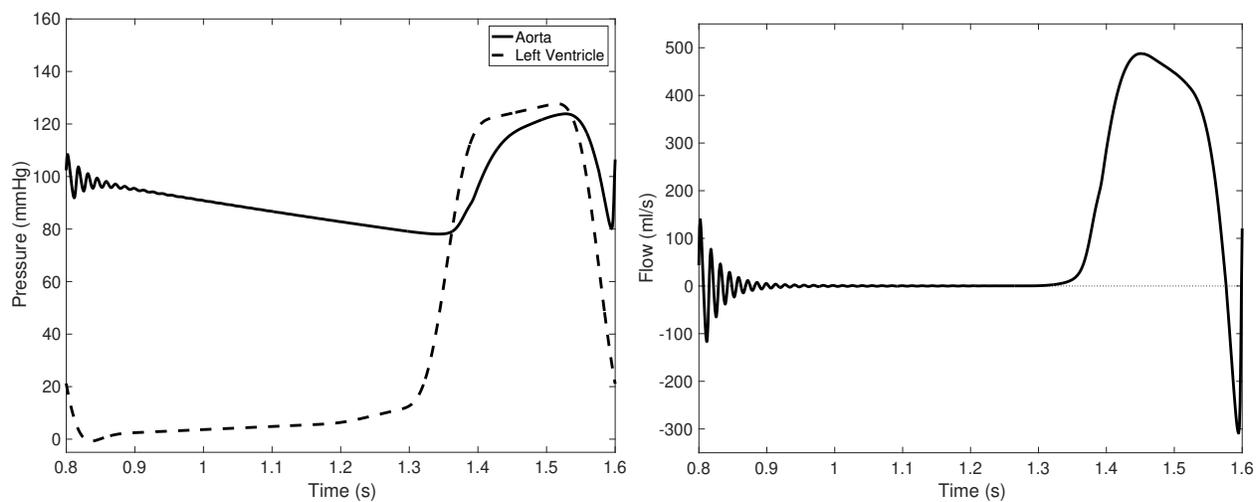

Figure 3: Pressure and flow waveforms on the 1.57d case from the second cardiac cycle.

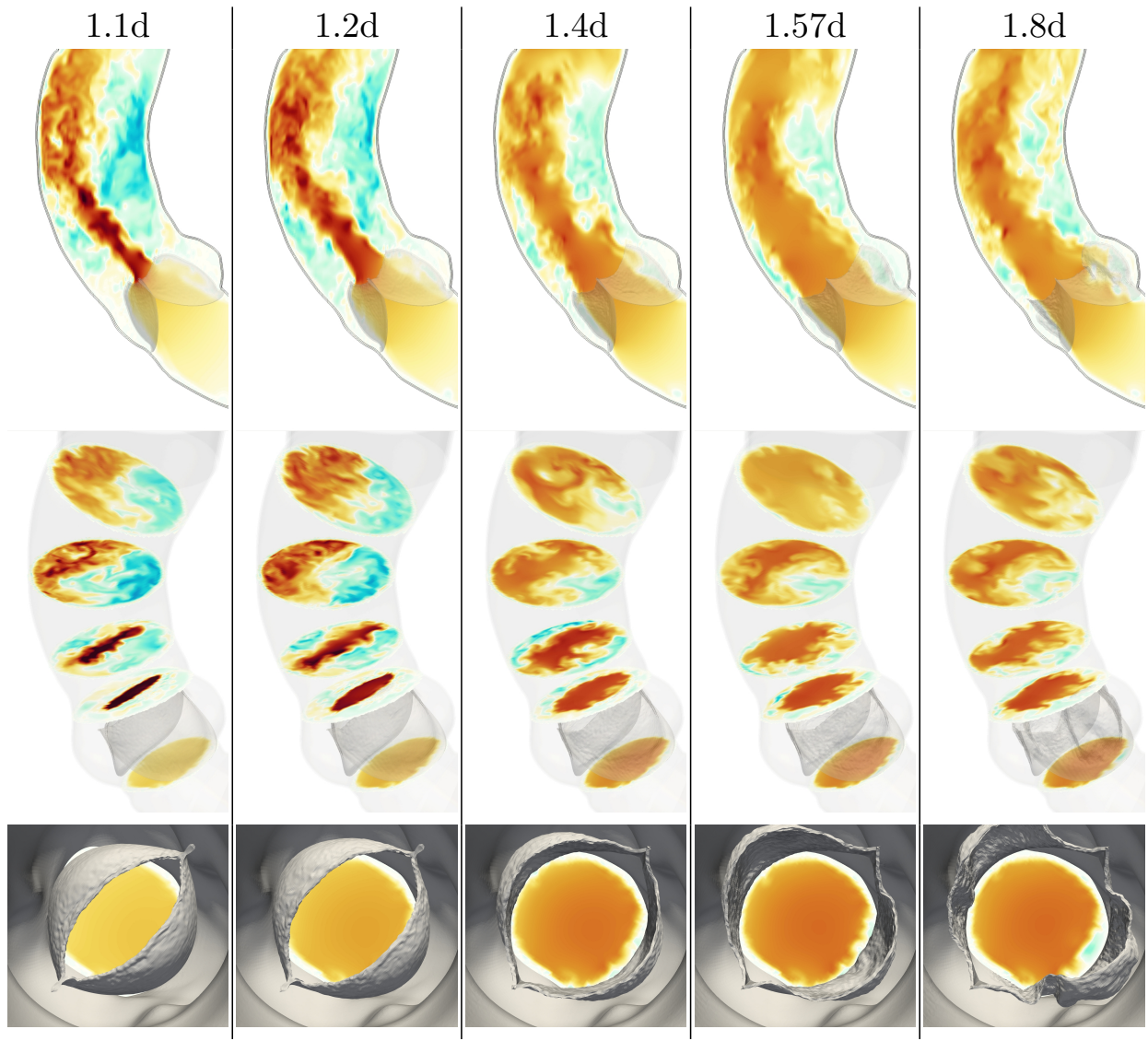

Figure 4: Flows in mid systole. The cases with free edge rest lengths of 1.1d and 1.2d show obvious stenosis, reduced orifice area, and narrowed jets of forward flow. The case with free edge rest length 1.4d has considerably less restriction; the cases with 1.57d and 1.8d have no apparent stenosis or restriction to forward flow.

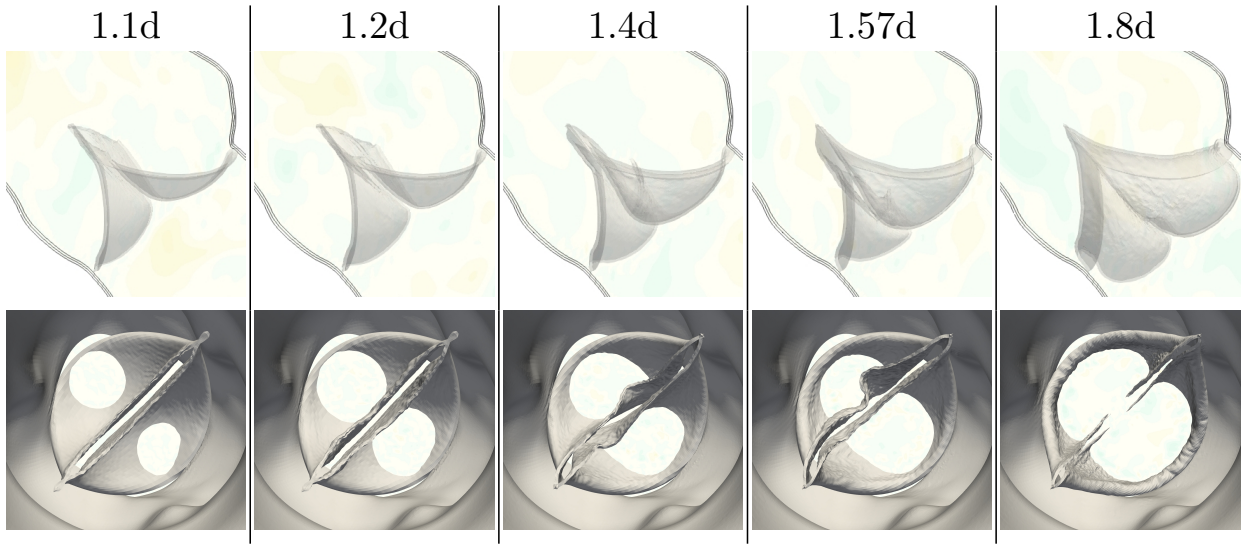

Figure 5: Flows in diastole. The valves are sealed with no regurgitation visible in all cases. Billow increases with increasing free edge length.

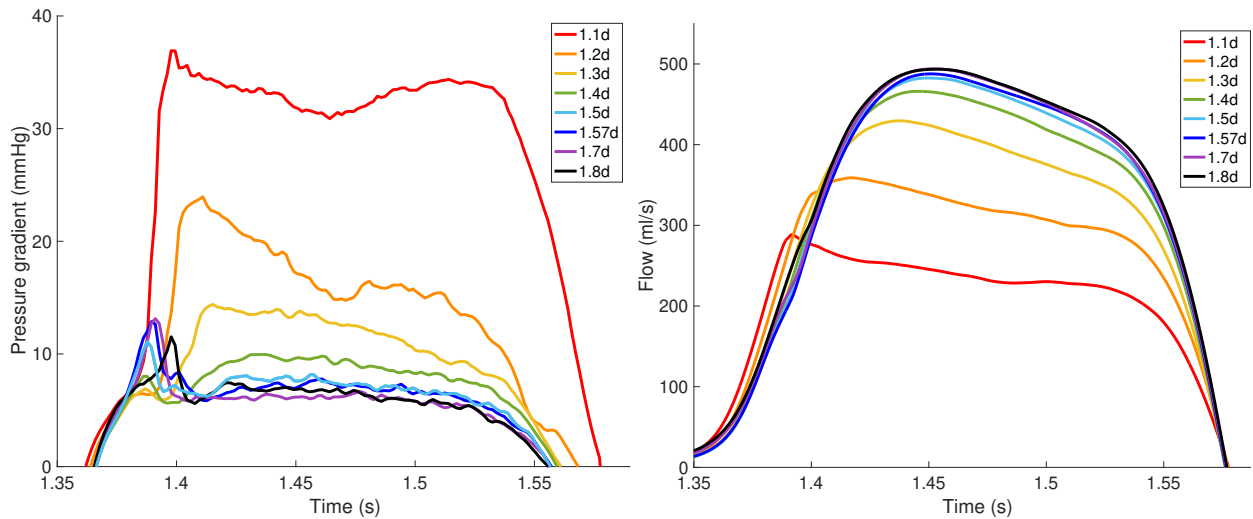

Figure 6: Pressure gradients and flow rates in systole compared across multiple cases. The cases with free edge length 1.3d or under show stenosis and reduced flow. The cases with free edge length ≥1.5d show pressure gradients under 10 mm and similar flow rates.

Systole

| case | orifice area | pressure gradient (mean) | stroke volume |
|------|------|------|------|
|  | cm$^2$ | mmHg | ml |
| 1.1d | 1.31 | 32.94 | 46.36 |
| 1.2d | 1.86 | 17.30 | 59.11 |
| 1.3d | 2.53 | 12.19 | 67.99 |
| 1.4d | 3.05 | 9.09 | 73.08 |
| 1.5d | 3.37 | 7.37 | 75.10 |
| 1.57d | 3.45 | 6.92 | 75.52 |
| 1.7d | 3.42 | 6.03 | 77.01 |
| 1.8d | 3.46 | 6.50 | 78.20 |

Diastole

| case | $G_h$ | $G_h/r$ | billow height | $E_h$ | $E_{h,min}$ | $E_h/G_h$ | $E_{h,min}/G_h$ | $F$ | $F/d$ |
|------|------|------|------|------|------|------|------|------|------|
|  | cm |  | cm | cm | cm |  |  | cm |  |
| 1.1d | 2.23 | 1.79 | -0.13 | 1.12 | 1.25 | 0.50 | 0.56 | 2.95 | 1.18 |
| 1.2d | 2.28 | 1.82 | -0.20 | 0.97 | 1.17 | 0.42 | 0.51 | 3.12 | 1.25 |
| 1.3d | 2.26 | 1.81 | -0.23 | 0.89 | 1.12 | 0.39 | 0.50 | 3.42 | 1.37 |
| 1.4d | 2.20 | 1.76 | -0.24 | 0.68 | 0.93 | 0.31 | 0.42 | 3.68 | 1.47 |
| 1.5d | 2.22 | 1.78 | -0.28 | 0.59 | 0.87 | 0.27 | 0.39 | 3.93 | 1.57 |
| 1.57d | 2.15 | 1.72 | -0.28 | 0.59 | 0.87 | 0.27 | 0.40 | 4.02 | 1.61 |
| 1.7d | 2.22 | 1.78 | -0.44 | 0.21 | 0.65 | 0.09 | 0.29 | 4.49 | 1.80 |
| 1.8d | 2.31 | 1.85 | -0.56 | -0.03 | 0.53 | -0.01 | 0.23 | 4.84 | 1.94 |

Table 1: Scalar metrics in systole and diastole.

| standard | area removed | height removed | height, area removed |
|----------|--------------|----------------|----------------------|

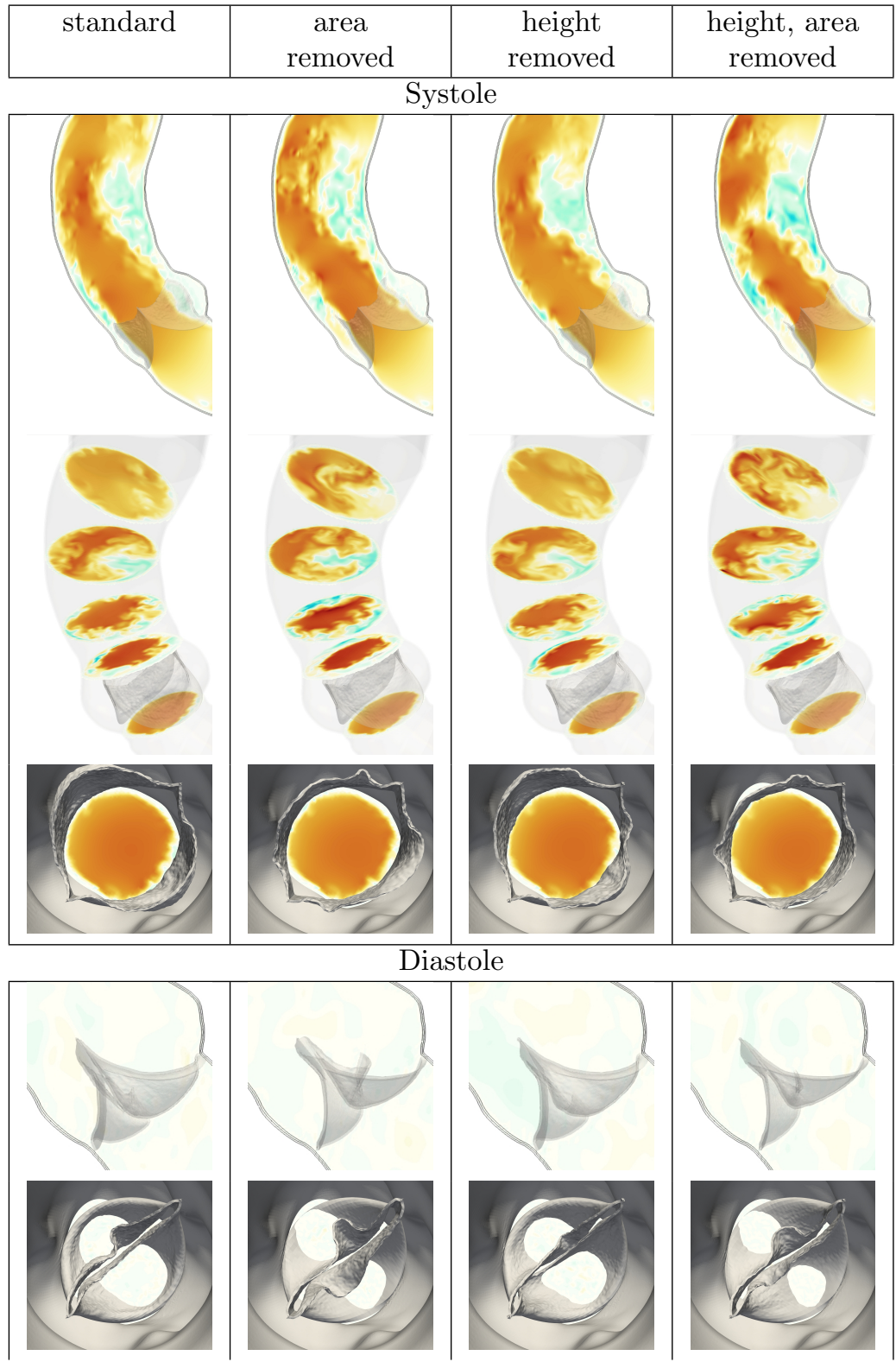

Supplementary Figure 1: Comparison of variations in gross morphology with similar free edge lengths.

Systole

| case | orifice area | pressure gradient (mean) | stroke volume |
|---|---|---|---|
| | cm$^2$ | mmHg | ml |
| Basic | 3.45 | 6.92 | 75.52 |
| Less area | 2.98 | 9.43 | 72.84 |
| Less height | 3.27 | 7.31 | 75.23 |
| Less area, height | 2.89 | 10.26 | 72.20 |

Diastole

| case | $G_h$ | $G_h/r$ | billow height | $E_h$ | $E_{h,min}$ | $E_h/G_h$ | $E_{h,min}/G_h$ | $F$ | $F/d$ |
|---|---|---|---|---|---|---|---|---|---|
| | cm | | cm | cm | cm | | | cm | |
| Basic | 2.15 | 1.72 | -0.28 | 0.59 | 0.87 | 0.27 | 0.40 | 4.02 | 1.61 |
| Less area | 2.22 | 1.78 | -0.17 | 0.86 | 1.03 | 0.39 | 0.46 | 4.02 | 1.61 |
| Less height | 1.88 | 1.50 | -0.23 | 0.30 | 0.54 | 0.16 | 0.29 | 4.29 | 1.71 |
| Less area, height | 1.94 | 1.55 | -0.12 | 0.63 | 0.74 | 0.32 | 0.38 | 4.07 | 1.63 |

Supplementary Table 1: Scalar metrics in systole and diastole.

Graphical abstract:

## Simulation-Based Design of Bicuspidization of the Aortic Valve

**Methods:**

Fluid-structure interaction simulations to evaluate and reduce stenosis in bicuspidization of the aortic valve

- One patient-specific aortic geometry
- Eight model bicuspid valves with free edge length from 1.1 to 1.8 times the annular diameter d

**Results:**

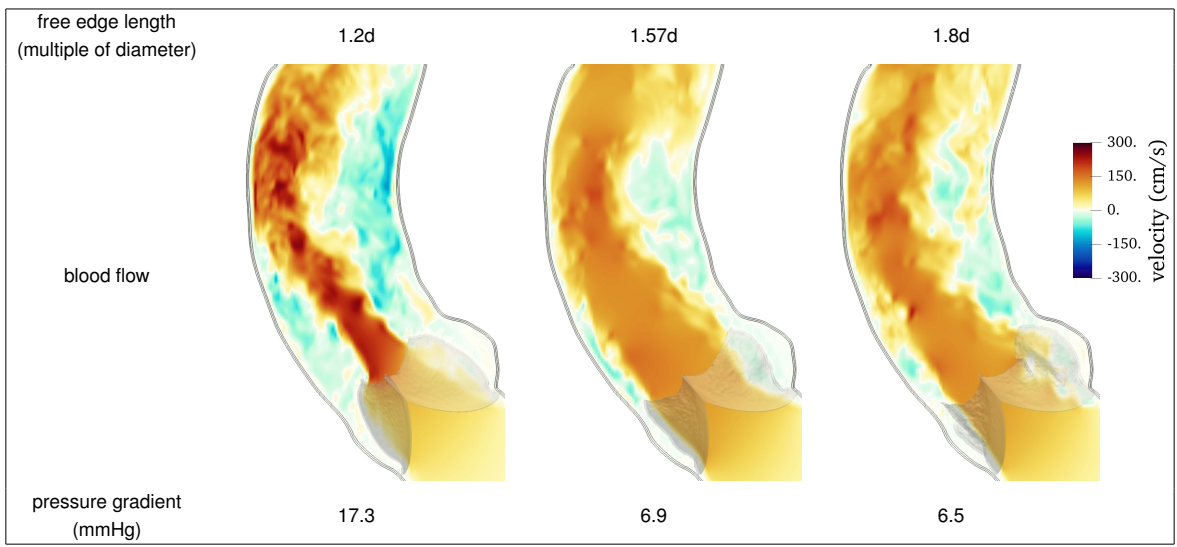

- Stenosis decreased with increasing free edge length.

**Implications:**

- Surgeons should consider increasing free edge length in bicuspidization of the aortic valve to reduce postoperative stenosis.